\documentclass[12pt]{article}
\usepackage{amsmath,amssymb,graphicx,array}
\usepackage{url,thophys,feynmp,emp}
\newenvironment{algorithm}%
 {\begin{list}{}%
   {\setlength{\leftmargin}{2em}%
    \setlength{\rightmargin}{2em}%
    \setlength{\itemindent}{1em}%
    \setlength{\listparindent}{0pt}%
    \settowidth{\labelwidth}{5em}%
    }}%
 {\end{list}}
\begin{document}
\begin{fmffile}{\jobname pics}
\fmfset{arrow_ang}{10}
\fmfset{curly_len}{2mm}
\fmfset{wiggly_len}{3mm}
\begin{empfile}
\title{%
  \parbox{\textwidth}{%
    \begin{flushright}\texttt{IKDA 2000/30}\end{flushright}}\\[\baselineskip]
  O'Mega: An~Optimizing Matrix~Element~Generator}
\author{Thorsten Ohl\\\hfil\\
  Darmstadt University of Technology\\
  Schlo\ss{}gartenstra\ss{}e 9, 64289 Darmstadt, Germany\\
  \texttt{ohl@hep.tu-darmstadt.de}}
\maketitle
\begin{abstract}
  I sketch the architecture of \textit{O'Mega}, a new
  optimizing compiler for tree amplitudes in quantum field theory.
  O'Mega generates the most efficient code currently available for
  scattering amplitudes for many polarized particles in the standard
  model.  A complete infrastructure for physics beyond the standard
  model is provided.
\end{abstract}

\section{Introduction}
Current and planned experiments in high energy physics can probe
processes with many tagged---potentially polarized---particles in the
final state.  The combinatorial explosion of the number of Feynman
diagrams contributing to scattering amplitudes for many external
particles calls for the development of more compact representations
that translate well to efficient and reliable numerical code.  In
gauge theories, strong numerical cancellations in a redundant
representation built from necessarily gauge dependent Feynman diagrams
lead to a loss of numerical precision, stressing further the need for
eliminating redundancies.

Due to the large number of processes that have to be studied in order
to unleash the potential of modern experiments, the construction of
these representations must be possible algorithmically on a computer
and should not require human ingenuity for each new application.

O'Mega~\cite{O'Mega} is a compiler for tree-level scattering
amplitudes that satisfies these requirements.  O'Mega is independent
of the target language and can support code in any programming
language for which a simple output module has been written.  To support a
physics model, O'Mega requires as input only the Feynman rules and the
relations among coupling constants.

Similar to earlier numerical approaches~\cite{ALPHA:1997,HELAC:2000},
O'Mega reduces the growth in calculational effort from a factorial of
the number of particles to an exponential.  The symbolic nature of
O'Mega, however, increases its flexibility.  Indeed, O'Mega can emulate
both~\cite{ALPHA:1997,HELAC:2000} and produces code that is empirically
at least twice as fast.

\section{1POWs And Keystones}

\emph{One Particle Off-shell Wave functions}~(1POWs)
are obtained from Greensfunctions by applying the LSZ reduction
formula to all but one line:
\begin{equation}
  W_{p_1,\ldots,p_n}^{q_1,\ldots,q_m}(x) =
     \Braket{\phi(q_1),\ldots,\phi(q_m);out|\Phi(x)
              |\phi(p_1),\ldots,\phi(p_n);in}\,.
\end{equation}
The 1POW
$W_{p}^{q,q'}(x)=\Braket{\phi(q),\phi(q');\text{out}|\Phi(x)|\phi(p);\text{in}}$ 
in lowest order of $\phi^3$-theory, is given---for illustration---by
\begin{equation}
  \parbox{26\unitlength}{%
    \fmfframe(2,3)(5,5){%
      \begin{fmfgraph*}(17,12)
       \fmflabel{$x$}{x}
       \fmflabel{$p$}{l}
       \fmflabel{$q$}{r1}
       \fmflabel{$q'$}{r2}
       \fmftop{x}
       \fmfleft{l,dl}
       \fmfright{r1,r2,dr}
       \fmf{plain}{l,v}
       \fmf{plain}{r1,v}
       \fmf{plain}{r2,v}
       \fmf{plain,tension=3}{x,v}
       \fmfblob{.4w}{v}
       \fmfdot{x}
      \end{fmfgraph*}}} =
  \parbox{26\unitlength}{%
    \fmfframe(2,3)(5,5){%
      \begin{fmfgraph*}(17,12)
       \fmflabel{$x$}{x}
       \fmflabel{$p$}{l}
       \fmflabel{$q$}{r1}
       \fmflabel{$q'$}{r2}
       \fmftop{x}
       \fmfleft{l,dl}
       \fmfright{r1,r2,dr}
       \fmf{plain}{l,v}
       \fmf{plain}{r1,vr,v}
       \fmf{plain}{r2,vr}
       \fmf{plain,tension=5}{x,v}
       \fmfdot{x}
      \end{fmfgraph*}}} +
  \parbox{26\unitlength}{%
    \fmfframe(2,3)(5,5){%
      \begin{fmfgraph*}(17,12)
       \fmflabel{$x$}{x}
       \fmflabel{$p$}{l}
       \fmflabel{$q$}{r1}
       \fmflabel{$q'$}{r2}
       \fmftop{x}
       \fmfleft{l,dl}
       \fmfright{r1,r2,dr}
       \fmf{plain}{l,vr,v}
       \fmf{plain}{r1,vr}
       \fmf{plain}{r2,v}
       \fmf{plain,tension=5}{x,v}
       \fmfdot{x}
      \end{fmfgraph*}}} +
  \parbox{26\unitlength}{%
    \fmfframe(2,3)(5,5){%
      \begin{fmfgraph*}(17,12)
       \fmflabel{$x$}{x}
       \fmflabel{$p$}{l}
       \fmflabel{$q$}{r1}
       \fmflabel{$q'$}{r2}
       \fmftop{x}
       \fmfleft{l,dl}
       \fmfright{r1,r2,dr}
       \fmf{plain}{l,vr}
       \fmf{plain,tension=0.5}{vr,v}
       \fmf{plain}{r1,v}
       \fmf{plain,rubout,tension=0.5}{r2,vr}
       \fmf{plain,tension=5}{x,v}
       \fmfdot{x}
      \end{fmfgraph*}}}
\end{equation}
At tree-level, the set of all 1POWs for a given set of external
momenta can be constructed recursively~\cite{HELAS}
\begin{equation}
\label{eq:recursive-1POW}
  \parbox{22\unitlength}{%
    \fmfframe(2,3)(2,1){%
      \begin{fmfgraph*}(17,15)
       \fmflabel{$x$}{x}
       \fmftop{x}
       \fmfbottomn{n}{6}
       \fmf{plain,tension=6}{x,n}
       \fmfv{d.sh=circle,d.f=empty,d.si=30pt,l=$n$,l.d=0}{n}
       \begin{fmffor}{i}{1}{1}{6}
         \fmf{plain}{n,n[i]}
       \end{fmffor}
      \end{fmfgraph*}}} = 
  \sum_{k+l=n}
  \parbox{32\unitlength}{%
    \fmfframe(2,3)(2,1){%
      \begin{fmfgraph*}(27,15)
       \fmflabel{$x$}{x}
       \fmftop{x}
       \fmfbottomn{n}{6}
       \fmf{plain,tension=8}{x,n}
       \fmf{plain,tension=4}{n,k}
       \fmf{plain,tension=4}{n,l}
       \fmfv{d.sh=circle,d.f=empty,d.si=20pt,l=$k$,l.d=0}{k}
       \fmfv{d.sh=circle,d.f=empty,d.si=20pt,l=$l$,l.d=0}{l}
       \fmffixed{(30pt,0pt)}{k,l}
       \begin{fmffor}{i}{1}{1}{4}
         \fmf{plain}{k,n[i]}
       \end{fmffor}
       \begin{fmffor}{i}{5}{1}{6}
         \fmf{plain}{l,n[i]}
       \end{fmffor}
       \fmfdot{n}
      \end{fmfgraph*}}}\,,
\end{equation}
where the sum extends over all partitions of the set of $n$~momenta.
For all quantum field theories, there are---well defined, but not
unique---sets of \emph{Keystones}~$K$~\cite{O'Mega} such that the sum
of tree Feynman diagrams can be expressed as a sparse sum of products
of 1POWs without double counting.  In a theory with only
cubic couplings this is expressed as
\begin{equation}
\label{eq:keystones}
  T = \sum_{i=1}^{F(n)} D_i =
      \sum_{k,l,m=1}^{P(n)}
        K^{3}_{f_kf_lf_m}(p_k,p_l,p_m)
        W_{f_k}(p_k)W_{f_l}(p_l)W_{f_m}(p_m)\,.
\end{equation}
The non-trivial problem of avoiding the
double counting of diagrams like (the circle denotes the keystone)
\begin{center}
   \begin{fmfgraph}(25,16)
     \fmfleftn{l}{3}
     \fmfrightn{r}{3}
     \fmf{plain}{l1,v4}
     \fmf{plain}{l2,v4}
     \fmf{plain}{l3,v4}
     \fmf{plain}{r1,v1}
     \fmf{plain}{r2,v1}
     \fmf{plain}{v1,v2}
     \fmf{plain}{r3,v2}
     \fmf{plain}{v2,v4}
     \fmfv{d.sh=circle,d.fill=empty,d.si=6thin}{v4}  
     \fmfdot{v1,v2}
   \end{fmfgraph}
   \qquad\qquad
   \begin{fmfgraph}(25,16)
     \fmfleftn{l}{3}
     \fmfrightn{r}{3}
     \fmf{plain}{l1,v4}
     \fmf{plain}{l2,v4}
     \fmf{plain}{l3,v4}
     \fmf{plain}{r1,v1}
     \fmf{plain}{r2,v1}
     \fmf{plain}{v1,v2}
     \fmf{plain}{r3,v2}
     \fmf{plain}{v2,v4}
     \fmfv{d.sh=circle,d.fill=empty,d.si=6thin}{v2}  
     \fmfdot{v1,v4}
   \end{fmfgraph}
\end{center} 
has been solved for general theories with vertices of arbitrary
degrees~\cite{O'Mega}.

The number of distinct momenta that can be formed from
$n$~external momenta is $P(n)=2^{n-1}-1$.  Therefore, the number of
tree 1POWs grows exponentially with the number of external particles
and not with a factorial, as the number of Feynman diagrams
$F(n)=(2n-5)!!=(2n-5)\cdot\ldots5\cdot3\cdot1$.  The
equations sketched in Eqs.~(\ref{eq:recursive-1POW})
and~(\ref{eq:keystones}) for cubic couplings can be generalized to
vertices of any order~\cite{O'Mega}.

Even for vector particles and to all orders in renormalized
perturbation theory, the 1POWs are `almost' physical objects and
satisfy simple Ward identities in unbroken gauge theories
\begin{equation}
    \frac{\partial}{\partial x_\mu}
    \Braket{\text{out}|A_\mu(x)|\text{in}}_{\text{amp.}} = 0
\end{equation}
and well as in spontaneously gauge theories 
\begin{equation}
    \frac{\partial}{\partial x_\mu}
    \Braket{\text{out}|W_\mu(x)|\text{in}}_{\text{amp.}} =
      \xi_W m_W \Braket{\text{out}|\phi_W(x)|\text{in}}_{\text{amp.}}
\end{equation}
in $R_\xi$-gauge.  The code for matrix elements can
optionally be instrumented by O'Mega with numerical checks of these
Ward identities for intermediate lines.

\section{Directed Acyclical Graphs}
The algebraic expression for the tree-level scattering amplitude in
terms of Feynman diagrams is itself a tree.  The much slower growth of
the set of 1POWs compared to the set of Feynman diagrams shows that this
representation is extremely redundant. In this case, \emph{Directed
Acyclical Graphs} (DAGs) provide a more efficient representation, as
illustrated by a trivial example
\begin{empcmds}
  vardef dag_coords =
    pair node[][]; node[1][1] = (.5w,.h);
    node[2][1] = (.3w,2/3h); node[2][2] = (.7w,2/3h);
    node[3][1] = (.2w,1/3h); node[3][2] = (.4w,1/3h);
    node[3][3] = (.6w,1/3h); node[3][4] = (.8w,1/3h);
    node[4][1] = (.5w,0/3h); node[4][2] = (.7w,0/3h);
  enddef;
  vardef dag_common =
    dag_coords;
    pickup pencircle scaled 1pt;
    label.rt (btex $\times$ etex, node[1][1]);
    draw node[1][1]--node[2][2];
    label.rt (btex $+$ etex, node[2][2]);
    draw node[2][2]--node[3][3];
    draw node[2][2]--node[3][4];
    label.rt (btex $\times$ etex, node[3][3]);
    draw node[3][3]--node[4][1];
    draw node[3][3]--node[4][2];
    label.rt (btex $\vphantom{b}c$ etex, node[3][4]);
    label.rt (btex $\vphantom{b}a$ etex, node[4][1]);
    label.rt (btex $\vphantom{b}b$ etex, node[4][2]);
    pickup pencircle scaled 3pt;
    pickup pencircle scaled 3pt;
    drawdot node[1][1];
    drawdot node[2][2];
    drawdot node[3][3];
  enddef;
\end{empcmds}
\begin{empdef}[dag](38,16)
  dag_common;
  pickup pencircle scaled 1pt;
  draw node[1][1]{(-1,-1)}..{(1,-1)}node[3][3];
\end{empdef}
\begin{empdef}[tree](38,16)
  dag_common;
  pickup pencircle scaled 1pt;
  label.rt (btex $\times$ etex, node[2][1]);
  draw node[1][1]--node[2][1];
  draw node[2][1]--node[3][1];
  draw node[2][1]--node[3][2];
  label.rt (btex $\vphantom{b}a$ etex, node[3][1]);
  label.rt (btex $\vphantom{b}b$ etex, node[3][2]);
  pickup pencircle scaled 3pt;
  drawdot node[2][1];
\end{empdef}
\begin{equation}
  ab (ab+c) =
  \parbox{28\unitlength}{\hfil\empuse{tree}\hfil}
    = \parbox{18\unitlength}{\hfil\empuse{dag}\hfil}\,,
\end{equation}
where one multiplication is saved.  The replacement of expression
trees by equivalent DAGs is part of the repertoire of optimizing
compilers, known as \emph{common subexpression elimination}.
Unfortunately, this approach fails for typical expressions appearing
in quantum field theory, because of the combinatorial growth of space
and time required to find an almost optimal factorization.

However, the recursive definition in Eq.~(\ref{eq:recursive-1POW}) allows to
construct the DAG of the 1POWs in Eq.~(\ref{eq:keystones})
\emph{directly}~\cite{O'Mega}, without having to construct and
factorize the Feynman diagrams explicitely.

As mentioned above, there is more than one consistent prescription for
constructing the set of keystones~\cite{O'Mega}.  The symbolic
expressions constructed by O'Mega contain the symbolic equivalents of
the numerical expressions computed by~\cite{ALPHA:1997} (maximally
symmetric keystones) and~\cite{HELAC:2000} (maximally asymmetric
keystones) as special cases.

\section{Algorithm}
By virtue of their recursive construction in
Eqs.~(\ref{eq:recursive-1POW}), tree-level 1POWs form a DAG and the
problem is to find the smallest DAG that corresponds to a given tree,
(i.\,e.~an given sum of Feynman diagrams).  O'Mega's algorithm
proceeds in four steps
\begin{algorithm}
  \item[Grow] starting from the external particles, build the tower of
    \emph{all} 1POWs up to a given height (the height
    is less than the number of external lines for asymmetrical
    keystones and less than half of that for symmetrical keystones)
    and translate it to the equivalent DAG~$D$.
  \item[Select] from $D$, determine \emph{all} possible
    \emph{flavored keystones} for the process under
    consideration and the 1POWs appearing in them.
  \item[Harvest] construct a sub-DAG $D^*\subseteq D$ consisting
    \emph{only} of nodes that contribute to the 1POWs
    appearing in the flavored keystones.
  \item[Calculate] multiply the 1POWs as specified by the keystones
    and sum the keystones.
\end{algorithm}
By construction, the resulting expression contains \emph{no} more
redundancies and can be translated to a numerical expression.  In
general, asymmetrical keystones create an expression that is smaller
by a few percent than the result from symmetrical keystones, but it
is not yet clear which approach produces the numerically more robust
results.

\section{Implementation}
The O'Mega compiler is implemented in O'Caml~\cite{O'Caml}, a
functional programming language of the ML family with a very
efficient, portable and freely available implementation, that can be
bootstrapped on all modern computers in a few minutes.

The powerful module system of O'Caml allows an efficient and concise
implementation of the DAGs for a specific physics model as a functor
application~\cite{O'Mega}.  This functor maps from the category of
trees to the category of DAGs and is applied to the set of trees
defined by the Feynman rules of any model under consideration.

The implementation is concise and efficient simultaneously by
exploiting the virtues of persistent data
structures~\cite{Okasaki:1998:book}.  Typically, the resources
consumed by O'Mega are only a small fraction of the resources required
by the compiler for the target language.

\begin{figure}
  \includegraphics[width=.9\columnwidth]{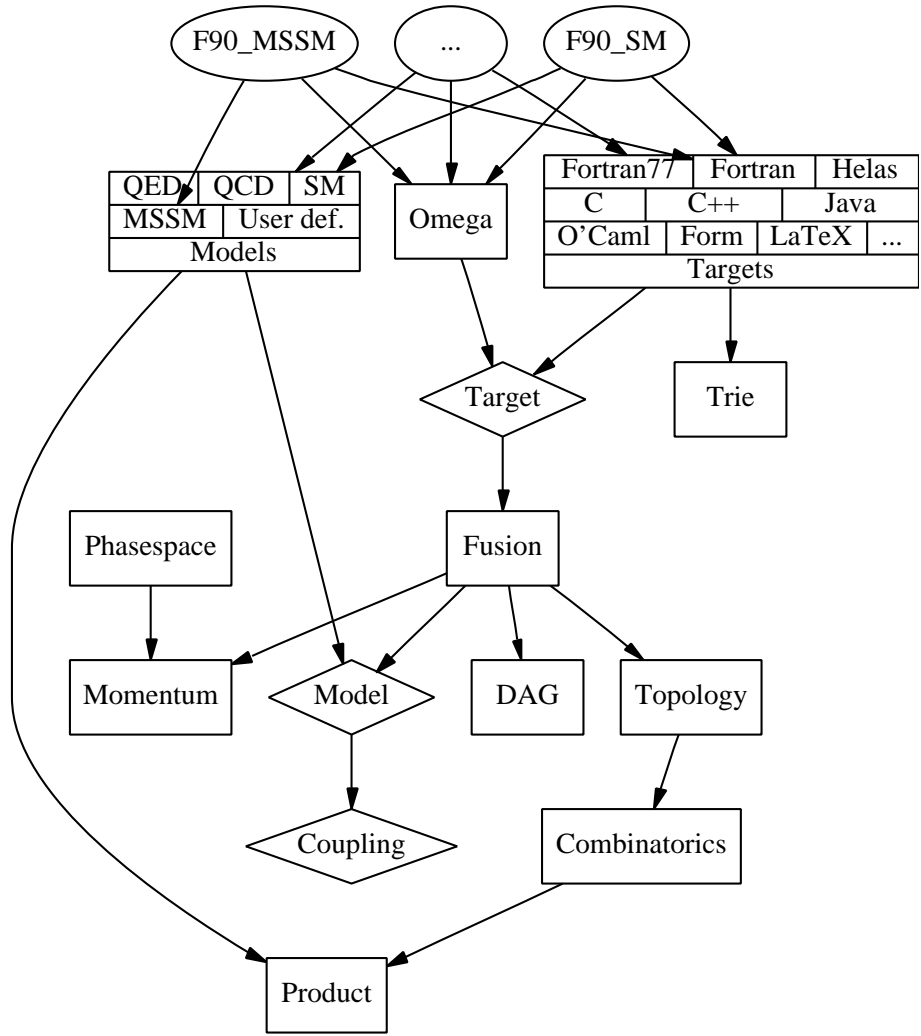}
  \caption{\label{fig:modules}%
    Module dependencies in O'Mega.  The diamond shaped nodes denote
    abstract signatures defining functor domains and co-domains.
    The rectangular boxes denote modules and functors, while oval
    boxes stand for example applications.}
\end{figure}
The module system of O'Caml has been used to make the combinatorial
core of O'Mega demonstrably independent from the specifics of both the
physics model and the target language~\cite{O'Mega}, as shown in
Figure~\ref{fig:modules}.  A Fortran90/95 backend has been realized
first, backends for C++ and Java will follow.  The complete
electroweak standard model has been implemented (the treatment of
interfering color amplitudes is still incomplete).
Majorana fermions, required
by supersymmetric field theories, are available
(using~\cite{Denner/etal:Majorana}) and the MSSM is in
preparation.

As mentioned above, the compilers for the target programming language are the
slowest step in the generation of executable code.  On the other hand,
the execution speed of the code is limited by non-trivial vertex
evaluations for vectors and spinors, which need $O(10)$ complex
multiplications.  Therefore, an \emph{O'Mega Virtual Machine} can
challenge native code and avoid compilations.

\section{Applications}
The code generated by the Fortran90/95 backend is the most efficient
code available for polarized scattering amplitudes for many particles.
The results have been compared with MADGRAPH~\cite{MADGRAPH:1994} for
many standard model processes and numerical agreement at the level
of~$10^{-11}$ has been found with double precision floating point
arithmetic.  O'Mega generated amplitudes are used in the omnipurpose
event generator generator WHIZARD~\cite{Kilian:WHIZARD}.  The first
complete experimental study of vector boson scattering in six fermion
production for linear collider physics~\cite{WWto6f} has been
facilitated by O'Mega and WHIZARD.

\subsection*{Acknowledgments}
I thank my collaborators Mauro Moretti and J\"urgen Reuter.  I thank
Wolfgang Kilian for valuable suggestions and for ``early adoption'' of
O'Mega.  This research is supported by the German
Bundesministerium f\"ur Bildung und Forschung (05\,HT9RDA) and
Deutsche Forschungsgemeinschaft (MA 676/6-1).


\end{empfile}
\end{fmffile}
\end{document}